\documentclass[12pt, leqno]{article}

\usepackage[margin=1.0in]{geometry}

\usepackage{amsmath,amsthm,amssymb,}
\usepackage{setspace}
\usepackage[dvipsnames]{xcolor}
\usepackage{wasysym, marvosym} 

\usepackage{color}
\usepackage{stmaryrd}

  \usepackage[pdftex,
  bookmarks = false,
  pdfstartview = FitBH,
  linktocpage = true,
  pagebackref = true,
  pdfhighlight = /O,
  colorlinks=true,
  linkcolor=blue,
  citecolor=blue,
  filecolor = blue,
  urlcolor = blue,
  menucolor = blue,
]{hyperref}

\usepackage[colorinlistoftodos]{todonotes}

\usepackage{fancyhdr}
\usepackage{physics}

\newcommand{\Marian}[2][]{\todo[color=green, #1]{#2}}

\usepackage{mathtools}

\usepackage[normalem]{ulem}

\usepackage{comment}

\title{Haag as a How-To Theorem}

\author{
\textsc{David Freeborn}\footnote{\href{mailto:dfreebor@uci.edu}{dfreebor@uci.edu} }\\
\textsc{Marian Gilton}\footnote{\href{mailto:marian.gilton@pitt.edu}{marian.gilton@pitt.edu}}\\
\textsc{Chris Mitsch}\footnote{\href{mailto:chris.mitsch@pitt.edu}{chris.mitsch@pitt.edu}}~\footnote{This work was produced through equal contribution from each author.}\\
}

\date{\today}

\begin{document}

\maketitle

\doublespacing

\begin{abstract}
    Haag's theorem is a classic no-go theorem. It rigorously demonstrates there is a logical problem with the interaction picture (IP), one of the most widely used modeling tools in quantum field theory (QFT). The significance of the theorem for the use of the IP in QFT has been the subject of long-running debate, focused around how ``worried'' we should be. In this paper, we argue for an alternative and opposite perspective on Haag's theorem, rejecting the `worry' framing in favor of emphasizing the no-go theorem's implications for model development.
\end{abstract}

\section{Introduction}

Haag's theorem is a classic no-go theorem. It rigorously demonstrates there is a logical problem with the interaction picture (IP), one of the most widely used modeling tools in quantum field theory (QFT). The significance of the theorem for the use of the IP in QFT has been the subject of long-running debate, focused around how ``worried'' we should be. One prominent slogan for the theorem is that `the IP doesn't exist.' Others argue that the IP remains sound despite Haag's theorem due to technicalities of its implementation.\footnote{See \cite{EarmanFraser2006implications, FraserDoreen2006Thesis} for crucial and seminal philosophical treatments of the significance of Haag's theorem. See \hyperlink{cite.FGM-Haag-framework_archiveversion}{[Freeborn, Gilton, and Mitsch, 2023]} \nocite{FGM-Haag-framework_archiveversion} \label{FGMarchivelink} for a recent and detailed set of mappings of the prominent viewpoints on the significance of Haag's theorem for the theoretical status of the IP as well as many other issues in philosophy of QFT.  } 
 Either we take the no-go theorem at face value and accept that one of contemporary physics's mainstay techniques is mathematically bankrupt, or else we accept that at least one of the fundamental assumptions of QFT must be at the very least bent out o shape if not outright abandoned in order to salve the IP. On this score, the debate has paralleled that around Bell's theorem, which seems to force us into a ``pick your poison'' trilemma. Call this the \textit{what-not} perspective on no-go theorems.

But this what-not perspective is not always correct. In this paper, we argue for an alternative and opposite perspective on Haag's theorem. On this how-to view, Haag's theorem is significant for the positive value it contributes for model building and for theory development.  
In this way, Haag's theorem is \textit{not} comparable to Bell's theorem. We therefore reject the antagonism of the `worry' framing in favor of emphasizing the no-go theorem's implications for model development. We call implications of the latter sort \textit{how-to} implications.

The argument proceeds by analogy with Arrow's theorem, extracting two lessons: (1) treating assumptions as sacrosanct could hinder theory development, and (2) ignoring an existing multiplicity of modeling purposes could conceal the model-development affordances of no-go theorems. We apply these lessons to Haag's theorem to highlight its how-to implications. We conclude that a how-to view of Haag's theorem better captures its significance.

\bigskip

\section{The how-to implications of Arrow's theorem}
To underscore the significance of Arrow's theorem, let's consider a real-world example: sustainable forest management.  In the mid-90s, the U.S.D.A. Forest Service was tasked with planning land use in parts of the Shoshone National Forest for a proposed oil and gas lease. In decades past, the Forest Service made decisions unilaterally \cite[603--4]{MARTIN1996603}. Yet, by the mid-90s, new legal frameworks and increasing public scrutiny had led to new decision-making processes under the rubric of ``sustainable forest management'' (SFM)\textemdash sustainable in the sense that it generates land management decisions that can be executed indefinitely, i.e., with less difficulty of implementation.

The Forest Service now needed to involve stakeholders (including the oil company, local industries, and government units) in its decision-making procedure. It identified several alternative plans \cite[617]{MARTIN1996603}. Naturally, the stakeholders disagreed about their preferred plan. The Forest Service needed a framework to implement a participatory planning that accounted for the stakeholders' varying preferences.

Happily, social choice theory provides just such a framework. Given its presumptive goal of ``combin[ing] individual preferences into collective choice'' \cite[vii]{mackay1980}, it seems well-positioned to render SFM for the  Forest Service. All the standard social choice ingredients are in place: several alternative land-use plans\textemdash call them \textit{alternatives}; multiple stakeholders whom we can model as \textit{voters} with individual \textit{preference schedules}: what the Forest Service needs is then an \textit{aggregation procedure} for (fairly) determining a collective choice.

Yet, social choice theory delivers troubling news for our Forest Service team in the form of \textit{Arrow's theorem}, a no-go theorem proven by Kenneth Arrow in 1951 \cite{Arrow_1950}. This theorem purports to show that, given at least three voters and two alternatives, \textit{no (fair) aggregation procedure exists}. The case at hand meets the conditions, so, we begrudgingly conclude, the Forest Service cannot fairly determine what to do with the land.

Nevertheless, the Forest Service has persisted in using participatory planning. Moreover, the processes implemented have led to more sustainable plans. This raises a question: if we can apparently just ignore it here, \textit{what is the significance of Arrow's theorem}?

As has long been recognized, the significance of Arrow's theorem resides not in its directly telling us which goals are impossible or, contrapositively, which theorems can be ignored by dint of a goal's achievement in practice. Instead, its significance has been in guiding the subsequent development of social choice theory. In other words, the how-to\textemdash rather than what-not\textemdash implications of Arrow's theorem are most significant.

\subsection{Malleability and Combinatorial Salvation}\label{Dardashti}

The first mistake invited by comparison to Bell's theorem concerns the sacrosanctity of no-go assumptions. It is generally agreed that Bell's theorem undercuts \textit{any} search for a (non-superdeterministic) local hidden variable theory of quantum phenomena (see \cite{Clauser_1978, Maudlin_2014}). Much of this assessment turns on the presumed sacrosanctity of its assumptions. 
Yet, to the degree that the assumptions are model-laden, this presumption is methodologically harmful.




We can see the value of treating assumptions as malleable (as apposed to sacrosanct) when we return to SFM and Arrow's theorem. Let's make our assumptions explicit. First, stakeholders are rational in the sense that their preferences exhibit a \textit{weak ordering}. Second, stakeholders may rank the alternatives in any possible order (\textit{unrestricted domain}). Third, what the Forest Service needs is an \textit{aggregation procedure}, a function from complete lists of stakeholder preferences to the set of \textit{social orderings}. A \textit{democratic} aggregation procedure requires that: social preferences are weakly ordered; if everyone prefers $A$ to $B$, then so do social preferences; no single voter can dictate the outcome; and modifications to the order of irrelevant alternatives cannot change the social preference profile. The Forest Service's goal is to find a democratic aggregation procedure; however, Arrow's theorem tells us there is no such procedure \cite{sep-arrows-theorem}.

While the goal may yet be impossible, the clear articulation of the assumptions and goal already points to the potential salvation. Picture the assumptions as faces of the dials of a combination lock, and ``unlocking'' it as  finding  an aggregation procedure. Arrow's theorem merely tells us that the \textit{chosen} combination doesn't open the lock. The Forest Service could simply change the dials by revising assumptions that don't capture their target system, suss out further background assumptions,\footnote{\label{Arrow_background_assumptions}Three are worth mentioning\textemdash aggregation procedures: are \textit{decisive} in the sense that they always generate a social ordering; are \textit{determinate} in the sense that the social ordering they generate are determined solely by their inputs; and deliver a ranking that is \textit{utility-maximizing}.} or even modify the specific goal they meant to achieve. Merely recognizing the number of possible combinations might suffice to quell Arrow-fueled worry.

This ``combinatorial'' understanding was fruitful in both revealing dials (i.e., seeing Arrow's assumptions as malleable), and leading to their turning, spurring social choice theory's development.\footnote{For this reason \cite[47]{Dardashti2021} calls no-go theorems ``methodological tools in theory development.''} For example, many (\cite{Hansson1973, Pazner1979, Sen1993, Breton_Weymark,Fleurbaey_2003}) have argued that several assumptions are too strong in any application. More radically, \cite{Balinski_Laraki} have shown that alternative aggregation procedures (scoring or grading rules) can escape Arrow's theorem. Finally, Arrow's theorem prompted \cite{Sen1999} to question the merits of the social choice framework itself.

In the case at hand, salvation was to be had. As for all proposals, the Forest Service had to prepare an Environmental Impact Statement, the legal framework for which invariably produces a natural ordering of alternatives \cite[614,617]{MARTIN1996603}. 
Since the Forest Service can always double-count itself to ensure an odd number of voters, an aggregation procedure exists (Condorcet winner). For another, the independence of irrelevant alternatives may be too strong, too: given the natural ordering on alternatives and a relatively equitable distribution of information, strategic voting is easier to detect and address. Again, Arrow's theorem is evaded.

What is clear is that the no-go theorem led to a blossoming of responses which have guided exploration of numerous methodological pathways forward in social choice theory. Notably, pessimism regarding the original goal of developing fair aggregation procedures has been the least popular path in this literature. Thus, while a no-go theorem \textit{does} show that a specific goal is impossible to attain given certain assumptions, practitioners, like the Forest Service, need not despair. While each no-go theorem deserves a native-tongue analysis, 
the combinatorial analogy highlights the pathways (dials) one could, in principle, explore to achieve the same goal.

So, what lesson does Arrow's theorem teach about no-go theorems that Bell's does not? We have seen that treating Arrow's assumptions as malleable opened new conceptual and mathematical paths to the original goal, relevant to social choice theory's applications. Arrow's theorem served as an impetus for substantial development \textit{toward} democratic aggregation procedures. A sacrosanct treatment of the assumptions would have harmed such theory development. In a slogan, the first lesson is this: viewing your assumptions as sacrosanct could endanger the development of your theory.

\subsection{Modeling Goal Iteration and SFM's Purpose}\label{sec:Parker}

Yet however far the first lesson might take us toward a no-go theorem's goal, Bell's theorem fails to teach us an even more critical lesson. With Bell's theorem, it is taken for granted that a would-be quantum theory's purpose is to generate probabilities corresponding to those predicted by standard theory. Bell's theorem seems to rule out the goal of a local hidden variable theory generating probabilities that violate Bell's inequality. There is a seemingly-inviolable connection between the \textit{general purpose} of a would-be quantum theory and the \textit{technical goal} for achieving that purpose, the latter of which is ruled out by the no-go theorem.
But this connection is not a general feature of no-go theorems and assuming as much obscures the way such theorems bridge concrete scientific modeling and abstract theory development.

The role of Arrow's theorem in SFM makes this second mistake apparent, too. Begin by asking: what makes social choice theory a good model of decision-making? Following \cite{Parker2020}, we assume that models are adequate when they serve their intended purpose. Importantly, adequacy does not turn only on representational merits; rather, models must also meet pragmatic constraints to be adequate. This can separate a model's achieving the specific goal ruled out by a no-go theorem\textemdash e.g., a democratic aggregation procedure\textemdash and its adequacy for a modeler's\textemdash e.g., the Forest Service\textemdash purpose. We can view adequacy as the finding of solutions within a multi-dimensional purpose space, indexed by the mathematical model and aspects of how the model is to be implemented. Some regions will be ruled out by model non-existence (e.g., no-go theorems), while others will be ruled out for impracticalities of implementation.


Recall that the Forest Service aims to deliver a sustainable plan. Effective participatory planning is expected to do this \cite[78]{kangas2006}. While many models are usable in-principle, the pragmatics of the case at hand constrain which are going to be adequate. In particular, the stakeholders are likely to withhold challenges to the decision only if it is \textit{perceived} to be fair and transparent. But this does not force the Forest Service to use social choice theory, let alone find a democratic aggregation procedure. That is, the Forest Service does \textit{not} need to satisfy the goal that Arrow's theorem rules out (modulo its assumptions) in order for a model to be adequate for its purpose. Indeed, if the stakeholders perceive the decision procedure as fair, does the Forest Service even need it to be decisive, determinate, or utility-maximizing? 

The geometry of the purpose space helps us understand the development and deployment of adequate models in SFM. First, the decision procedures do not need to be democratic. Second, since the legal framework generates alternatives with a natural order, the original assumptions of Arrow's theorem were too restrictive. Third, the Forest Service may not require even the most basic of properties of the aggregation procedure, depending on the circumstances. The purpose space has room for many more adequate models, beyond those excluded by Arrow's Theorem.

Importantly, some of this space has been filled in with decision procedures whose models are not strictly captured by social choice theory, (for instance, see \cite{SheppardMeitner2005}). Conversely, compilation of knowledge about various models' satisfaction of Arrow's theorem's goal (and like theorems) has facilitated more perspicuous selection of models in application (e.g., see \cite[89]{kangas2006}). Thus, Arrow's theorem has served as a link between abstract theory development and concrete model deployment. 

Elucidating these ``how-to'' implications paints a richer picture of the significance of no-go theorems. In particular, the goals of no-go theorems often simplify the purposes for which models are built to the point that the former are neither necessary nor sufficient for the latter's adequacy. Nevertheless, insofar as a no-go theorem's goal can be modified so as to better capture a modeler's purpose, it affords the development of better models. Conversely, insofar as a no-go theorem guides the selection of models more likely to be adequate for purpose, it also affords the development of more perspicuous and efficient modeling practices. In a slogan, the second lesson is this: recognizing the liminal space between no-go goals and modeling purposes can encourage model development and improve model deployment.

\section{How-to for Haag's theorem}\label{sec:howtohaag}

We now turn to applying our two lessons from Arrow's theorem to Haag's theorem. First, the assumptions of Haag's theorem are not sacrosanct. Second, Haag's theorem can still be epistemically valuable outside of its original modeling context. Taken together, these two lessons constitute a novel position on the significance of Haag's theorem. On this ``how-to'' view, Haag's theorem is scientifically valuable for its \textit{positive contributions to} the development of mainstay techniques for calculating scattering amplitudes, and not, as the what-not view would have it, as an inconvenient impediment for such techniques.

\subsection{The Malleability of Haag's Theorem's Assumptions}
The central purpose of scattering theory in QFT is to generate predictions about interactions that we can compare against observations, e.g. for scattering experiments. This involves specifying how the interacting system behaves and ``evolving'' the inputs to the outputs. Unfortunately, the complexity of the relevant interacting systems presents numerous mathematical obstacles to generating exact predictions. Various modeling techniques pursue the purpose of scattering with different concrete goals. A sizable portion of theoretical particle physics proceeds, broadly, with the following modeling goal: idealize the ``in'' and ``out'' multi-particle states (resp. $\langle \beta \lvert$, $\rvert \alpha \rangle$) as free of interactions, occurring at $-\infty$ and $+\infty$, respectively; let an interpolating field between free ``in'' and ``out'' states handle the interaction; and consider only the amplitudes of particular scatterings rather than the full dynamics of interaction. Call this the ``interpolating field'' approach to scattering theory. 


Worriers sometimes suppose Haag's theorem undermines any such use of an interpolating field approach to rigorously derive (non-trivial) scattering amplitudes (e.g., \cite{Redhead1990-REDAPL,sklar2000theory}). 
Such reactions betray an understanding of the theorem's assumptions as sacrosanct and model-independent. There are three main classes of assumptions (cf. \cite[253-268]{duncan2012conceptual}). The \emph{state axioms} require that: (S1) states form a separable Hilbert space carrying a unitary representation of the Poincar{'e} group; (S2) the infinitesimal generators of the translation subgroup have a spectrum restricted to the forward light-cone; and (S3) there is a unique vacuum state. The \emph{field axioms} require that: (F1) (interacting/interpolating) fields are represented as unbounded operators defined on a dense subset of a Hilbert space such that their (finite) products are also in the dense subset; (F2) the fields transform under the unitary representation of the Poincar{'e} group as expected; (F3) the commutator vanishes for fields defined on space-like separated regions; (F4) all states can be approximated arbitrarily well by linear combinations of field operators applied to the vacuum. The \emph{duality axioms} require that: (D1) the interpolating field is given in the interaction picture; and (D2) the ``in'' and ``out'' state spaces are unitarily equivalent.



Focus on (D1), which tells us how systems time-evolve. Quantum mechanics usually works within the Schr{\"o}dinger picture, where states, not operators, time-evolve. Relativistic symmetries make this difficult in QFT. Equivalently, we could let operators time-evolve but not states (Heisenberg picture). This works better with relativistic symmetries, making it preferable in QFT. However, calculating in the Heisenberg picture is difficult. Finally (and equivalently), we can use the interaction picture (IP), in which we split fields into their free and interacting components ($H_0$ and $V=H-H_0$, resp.), which capitalizes on the strength of both.


This works as follows. The scattering amplitude of a multi-particle state $\rvert \alpha \rangle$ time-evolving to $\langle \beta \lvert$ by some $\Phi(x)$ has the form
\begin{displaymath}
\underbrace{\langle \beta \lvert e^{iHt} e^{-iH_0t}}_\text{$V$-dep. out-state} \underbrace{e^{iH_0t} \Phi(x) e^{-iH_0t}}_\text{$H_0$-dep. operator} \underbrace{e^{iH_0t} e^{-iHt} \rvert \alpha \rangle}_\text{$V$-dep. in-state}.
\end{displaymath}
 The IP's neat trick involves two steps. First, the interaction-picture field $\Phi(x,t)=e^{iH_0t} \Phi(x) e^{-iH_0t}$ evolves according to the free dynamics, while the states carry the interaction. But by (F4), we can approximate any state by application of field operators to the vacuum, so $\rvert \alpha \rangle$ actually has the form $e^{iVt}\Psi_{\alpha}(x)e^{-iVt} \rvert 0 \rangle_V$, where $\Psi$ is the field representing interactions, $\Psi_{\alpha}(x)$ is a polynomial of $\Psi$-field operators approximating $\rvert \alpha \rangle$, and $\rvert 0 \rangle_V$ is the interacting vacuum; call this the Heisenberg field. The second step uses the fact that the various pictures coincide at $t=0$. This allows us to relate the interacting ($V$) and free ($H_0$) states by a unitary transformation. The existence of such a transformation entails that the free and interacting states occupy the same state space. This ostensibly accomplishes the goal of the interpolating field approach.

Unfortunately, the IP's neat trick undermines its use when working with a unitary representation of the Poincar{\'e} group. Our two fields must coincide and are \textit{both} free. This is the essence of Haag's theorem, and it directly contradicts our specified goal.


Yet on inspection, (D1) does not seem to be model-independent. Rather, it is a way to (i) simplify interpolating field dynamics and (ii) ensure the coincidence of the ``in'' and ``out'' state spaces with the full state space. But (i) is obviously not necessary, and (ii) might be achieved by other means. Haag-Ruelle scattering does just this by weakening (D1): now the interpolating field is given by a Heisenberg field for the full, interacting-field. Then one can rigorously construct ``in'' and ``out'' state spaces using products of full field operators and, thereby, derive our (non-trivial) scattering amplitudes (\textit{Haag Asymptotic Theorem}). Assuming the \textit{Haag Asymptotic Condition}, wherein amplitudes built from the interpolating field converge to those built from  free in-out fields, we can even recover S-matrix amplitudes in terms of expectation values of time-ordered Heisenberg fields (i.e., the LSZ theory).

LSZ is thus a natural iteration of the IP because free in-out states are still related by an interpolating field. It differs in that full, interacting-field states do not need to converge to free in-out states. 
Instead, they need only generate the same scattering amplitudes (i.e., preserve inner product structure). Consequently, while the S-matrix is unique, the \emph{interpolating field} is not. Thus, the how-to view emphasizes Haag's theorem's role in guiding the iterative modeling process from the naive IP to the sophisticated LSZ formalism for accomplishing the specific goal of the interpolating field approach. This is exactly opposite to the standard what-not view, which takes Haag's theorem to primarily be a stumbling block to achieving that modeling goal. 

Nevertheless, one might still worry that Haag's theorem should threaten perturbative calculations, which use the IP. However, in practice we \emph{also} fully regularize our theories which, by breaking Poincar{\'e} covariance or otherwise ensuring a finite-dimensional state space, evades the theorem (see \cite{duncan2012conceptual, miller2018haag}). What about when we remove cutoff dependence during renormalization, and purportedly restore Poincar{\'e} covariance? Even then, we only satisfy ourselves with the Poincar{\'e} covariance of the subset of terms used in the calculation, which does \textit{not} restore Poincar{\'e} covariance in the (S1) and (F2) sense. Thus, lesson one applies when the IP is deployed: Haag's theorem's assumptions are not sacrosanct.

Since the assumptions are not sacrosanct, we should reject the what-not view of Haag's theorem. 
By adjusting the duality assumptions, Haag-Ruelle and LSZ demonstrate the interpolating field approach is a coherent modeling strategy. Likewise, tracking the theorem's assumptions across IP calculations reveals the shifted meaning of Poincar{\'e} covariance. On the what-not view, the central question is whether Haag's theorem is a source of serious worry that our modeling goal is not achievable since either the stated goal or at least one of the sacrosanct assumptions must be sacrificed in order to evade the theorem and its dire consequences. When we apply lesson one of the how-to view, we arrive at the opposite assessment: Haag's theorem \textit{helps} us in the iterative modeling practice by suggesting opportunities for changing modeling assumptions. This is cause for optimism rather than worry since our assumptions were never meant to be sacrosanct, set in stone for all eternity. 




\subsection{Iterations On QFT's Purposes and Modeling Goals }


Outside the context of the specific goal associated with a no-go theorem, one might have thought the theorem has no further how-to implications to offer for the pursuit of other goals or purposes. But Arrow's theorem showed us that this is not always the case. As we saw in section \ref{sec:Parker}, the how-to implications of Arrow's theorem encompass both concrete model deployment and more abstract theory development.  In this section, we argue for the analogous lesson in the case of Haag's theorem. 

Recall the central purpose of the area of QFT research to which Haag's theorem most immediately applies, namely, generating predictions about interactions to which we can compare observations. We can pursue this same purpose with alternative technical goals to the interpolating field approach to accomplish. One such radical change of goal is argued for by \cite{kastner_directaction}.  Kastner suggests that the central problem of QFT, leading to Haag's theorem, is \textit{precisely} the assumption that interpolating fields mediate between in-out states. 
Instead, Kastner proposes the elimination of these field states by moving to a direct-action theory. Curiously, a related observation fuels Seidewitz's rehabilitation of the IP \cite{seidewitz2017avoiding}. However, his version of the IP does not rely on the interpolating field approach \textit{per se} since his field evolves according to a time-invariant parameter encoding paths in spacetime. 
His proposal recovers the perturbative expansions for scattering amplitudes term-by-term. Both  Kastner's and Seidewitz's proposals for substantive theory development make explicit use of Haag's theorem in identifying and motivating their respective avenues of research. Though both have given up on the original interpolating field goal in their own way, the no-go theorem still has methodological value to import from this model-building arena to the new arena of abstract theory development.  

Or perhaps we are seeking a successor theory by building upon aspects of QFT. The inability to generate usable predictions is not necessarily damning for this purpose. For instance, we might need to understand the limitations that Haag's theorem places on non-perturbative renormalization techniques. Here we are concerned with precisely the models susceptible to Haag's theorem. For example, it has been shown that the ambiguities in the full perturbative expansion arising from Haag's theorem can be reduced to a single one \cite[8-10]{maiezza2020haag}. Alternatively, suppose we wish to countenance gravity in QFT. To do this, \cite[2]{hollands2010axiomatic} suggest, it is ``essential'' to abandon the interpolating field approach because ``there will not, in general, even be an asymptotic notion of particle states.'' Thus not only is the goal of the last section insufficient, even the broader purpose of modeling the behavior of particle-like objects is not held fixed. Despite this, the new framework provides insight into the non-analytic dependencies related to Haag's theorem \cite[37]{hollands2010axiomatic}.
In each of these pursuits, Haag's theorem helpfully organizes the new insights; this organization can itself have how-to implications for the purposes of theory development. 

Thus, as with Arrow's theorem, the how-to reading provides a richer picture of the connection between the mathematical result and its implications for modeling. Haag's theorem does not capture a singular purpose as directly as does Bell's theorem. The implications of Haag's theorem are no less significant for this; but they have a more \emph{methodological}, and less \emph{metaphysical} character. Haag's theorem plays a critical role in the task of QFT model construction. Unlike Bell's theorem, there is enough room between the specific goal Haag's theorem rules out and relevant broader purposes that it can serve as a useful organizational and exploratory fulcrum for relating various goals and purposes. We contend that to ignore this how-to potential is to ignore the very things that make Haag's theorem significant.

\section{Conclusion }

Discussions of Haag's theorem in the philosophy literature tend to fall into one of two camps. There are those who take the theorem at face-value: it says that we have no means of modeling interactions, or that conventional QFT derives scattering amplitudes from unsound methods. In response, a second camp emphasizes the simple logic behind how some calculational techniques of conventional QFT at some point negate at least one of the key assumptions needed to prove Haag's theorem: the `worry' is thus assuaged. We have argued that this dialectic suffers from a misunderstanding of the epistemic value of a no-go theorem such as Haag's. The epistemic value of a no-go theorem does not boil down to a straightforward adjudication of the possibility of achieving a specified modeling goal. 
Rather, many no-go theorems are epistemically valuable for their how-to implications. 

By drawing attention to a no-go theorem's how-to implications, we do not mean to detract from their logical identity as no-go's. The theorems under consideration are all recognized as yielding a technical result on the basis of a valid argument. Thus, this highly specific epistemic value of validity is not in question. But once we have such a proven no-go theorem in hand,  there is the subsequent question of what further  implications we should draw from it. 
We have argued that an important and overlooked 
epistemic value of  no-go theorems such as Haag's arises from the concrete and systematic guidance they provide to the modeler for generating alternative ways to achieve her goal, as well as in the bridge these theorems provide between practical model-building and mathematical theory development.

\renewcommand\refname{Bibliography}
\bibliographystyle{apalike}
\bibliography{HaagBibliography.bib}{}

\end{document}